\documentclass[aps,prl,reprint,amsmath,amssymb,superscriptaddress,floatfix,twocolumn]{revtex4-2}
\pdfoutput=1

\usepackage{amssymb, bm, amsmath, graphicx, physymb}
\usepackage[colorlinks=true,linktocpage=true,linkcolor=blue,citecolor=blue]{hyperref}
\usepackage[usenames,dvipsnames]{color}
\usepackage{slashed}
\usepackage[utf8]{inputenc}
\usepackage{enumerate}
\usepackage{amsfonts}
\usepackage{mathtools}
\usepackage{latexsym}
\usepackage{hyperref}
\usepackage{epstopdf} 
\usepackage{bbm}
\usepackage{soul}
\usepackage{float} 
\usepackage{titlesec}
\hypersetup{breaklinks=true}

\titleformat{\section}[runin]
{\normalfont\itshape}{\thesection}{1em}{}{} 
\titlespacing{\section}{0pt}{0.1\baselineskip}{\baselineskip}

\titleformat{\subsection}[runin]
{\normalfont\itshape}{\thesubsection}{1em}{}{} 
\titlespacing{\subsection}{0pt}{0.1\baselineskip}{\baselineskip}

\titleformat{\subsubsection}[runin]
{\normalfont\itshape}{\thesubsection}{1em}{}{} 
\titlespacing{\subsubsection}{0pt}{0.1\baselineskip}{\baselineskip}

\begin{document}
%

%
\title{Signatures of Jet Drift in QGP Hard Probe Observables}


\author{Joseph Bahder}
\email[Email: ]{jbahder@nmsu.edu}
\affiliation{Department of Physics, New Mexico State University, Las Cruces, NM 88003, USA}

\author{Hasan Rahman}
\email[Email: ]{hrrahman@nmsu.edu}
\affiliation{Department of Physics, New Mexico State University, Las Cruces, NM 88003, USA}
\author{Matthew D. Sievert}
\email[Email: ]{msievert@nmsu.edu}
\affiliation{Department of Physics, New Mexico State University, Las Cruces, NM 88003, USA}
\author{Ivan Vitev}
\email[Email: ]{ivitev@lanl.gov}
\affiliation{Theoretical Division, Los Alamos National Laboratory, Los Alamos, NM 87545, USA}

%
\begin{abstract}
    Hard probe tomography of the quark-gluon plasma (QGP) in heavy ion collisions has long been a preeminent goal of the high-energy nuclear physics program. In service of this goal, the isotropic modification of jets and high-energy hadrons has been studied in great detail at the leading-power (eikonal) level, with effects originating from sub-eikonal $\mathcal{O}(\mu/E)$ anisotropic interactions  presumed to be small. We present the first investigation of sub-eikonal, collective-flow-induced asymmetric jet broadening (jet drift) in event-by-event $\sqrt{s}=5.02$ TeV PbPb collisions at the Large Hadron Collider using the new Anisotropic Parton Evolution (APE) computational framework. We show that jet drift imparts a sizeable and measureable enhancement of elliptic flow ($v_2$) and increases the mean acoplanarity for low and intermediate energy particles ($p_T < 10$ GeV). Importantly, these novel modifications to hard probe observables are shown to survive averaging over events and collision geometry. They couple to the collective flow of the medium seen by the jet and encode information about the QGP dynamics inaccessible to studies considering only isotropic, eikonal level effects.
\end{abstract}
%

\date{\today}
\maketitle






%
\section{Introduction.\label{sec:introduction}}
%
\begin{figure}[t]
    \centering
    \includegraphics[width=0.5\textwidth]{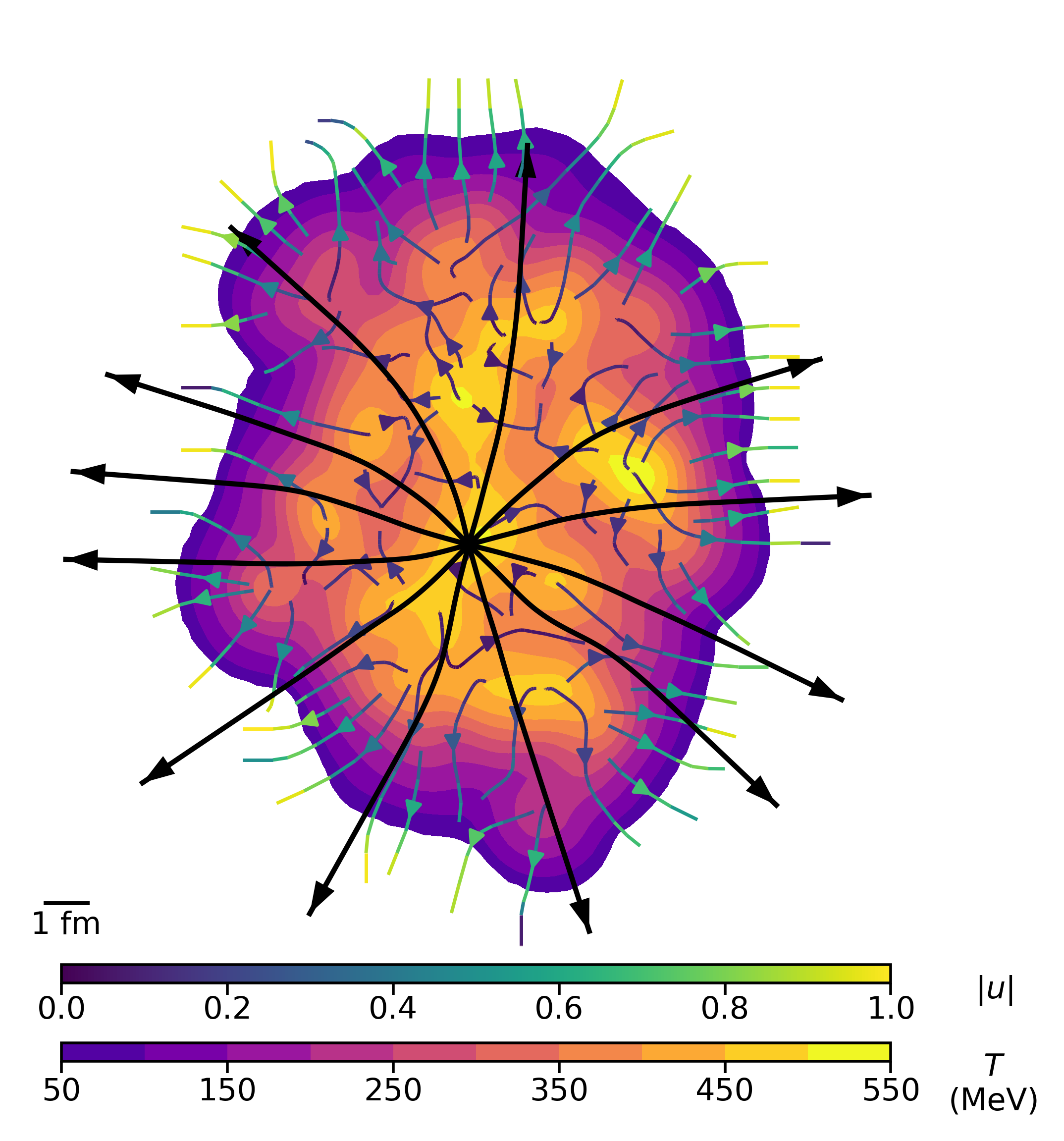}
    \caption{APE computed trajectories (black arrows) for an azimuthally uniform distribution of gluons originating at a hot spot in a mid-central $\sqrt{s}_{NN} = 5.02$ TeV PbPb event geometry. Note the coupling to the elliptic flow of the medium, attracting partons to the event plane. Parameters selected for illustrative purposes. \label{fig:sample_trajectory}}
\end{figure}
Hard probes are a cornerstone in the  investigation of the novel form of nuclear matter  -- the quark-gluon plasma -- produced in heavy ion collisions at the Relativistic Heavy Ion Collider (RHIC) and the Large Hadron Collider (LHC) (see \cite{Connors:2017ptx} for a review). Following initial and acclaimed success~\cite{Vitev:2002pf}, progress into hard probe tomography, especially recently,  has been limited by the inherent complexity of the problem. The standard measurement of the suppression of energetic particles and jets, the nuclear modification factor
\begin{align} \label{e:RAA}
    R_{AA}(p_T) & = \frac{dN^{AA}}{dyd^2p_T} \bigg{/} \langle n_{coll} \rangle\frac{dN^{pp}}{dyd^2p_T},
\end{align}
where $\langle n_{coll} \rangle$ is the number of binary collisions, is a robust quantity which can be reproduced by a wide variety of pQCD calculations and other models 
\cite{Noronha-Hostler:2016eow, Molnar:2013eqa, Zhao:2021vmu, Zhang:2013oca, Kopeliovich:2012sc, Andres:2019eus} (see \cite{CMS:2024krd} for a direct multi-model comparison). 
Other traditional hard probe modifications like isotropic transverse momentum broadening 
measured in
two-particle away-side correlations are also poorly discriminating, as they are complicated by competing background effects like energy-resolution smearing and medium response \cite{Connors:2017ptx, ALICE:2023qve}.
In contrast, there are few, if any, background effects that can mimic the anisotropic coupling of energetic partons to preferred directions in the medium.  Anisotropic modifications which go beyond the leading-power eikonal limit are, thus, more discriminating with respect the microscopic Quantum Chromodynamics (QCD) interactions in matter than their isotropic counterparts and encode unique information about the quark-gluon plasma (QGP). 

In previous work \cite{Sadofyev:2021ohn},  as part of the effort to develop the full  theoretical framework of QCD dynamics in matter, the authors have shown that the momentum broadening of a hard parton has an $\mathcal{O}(\mu/E)$ anisotropic component, $\mu$ being a medium scale and $E$ the parton energy,  which couples to the local collective flow velocity $\vec{u}$ in a thermal QGP. On average, this results in a net deflection of the parton in the direction of $\vec{u}$: the so called ``jet drift" effect.  The aggregation of the locally anisotropic jet drift  over the path length creates a global correlation between the inclusive azimuthal distribution of hard particles and the geometric anisotropy of the medium (the eccentricity $\varepsilon_n$).  This enhances the corresponding anisotropic flow $v_n^{hard}$ of hard particles \cite{Antiporda:2021hpk}, defined by standard Fourier decomposition as
\begin{align}
    \frac{R_{AA}(p_T, \phi)}{R_{AA}(p_T)} & \equiv  
    1 + 2 \sum_{n = 1}^\infty v_n^{hard} \, \cos [n (\phi - \psi_n^{hard} (p_T))] \, , 
\end{align}
where $\psi_n^{hard}$ is the angle of the $n^{th}$ harmonic hard particle event plane \cite{Noronha-Hostler:2016eow}. We emphasize that jet drift represents an additional level of coupling to the medium geometry, distinct from the sensitivity to the path length dependence, although jet drift does nontrivially mix with these effects.
However, $ \mu/E \ll 1$ at large parton energy $E$ suggests that modifications of hard probe observables might be small. Furthermore, one may reasonably question whether anisotropies due to the medium velocity field survive averaging over fluctuating events.  Thus, it is imperative to investigate this question on an event-by-event level. We note that power-suppressed (``sub-eikonal'') corrections like jet drift can modify every part of the jet production cross section: the hard scattering, the medium matrix elements, and the hadronization process.  Jet drift, which selects on locally \textit{asymmetric} interactions that can only come from the medium, therefore provides a first window into effects that go beyond the well-understood, leading-power formalism.

In this work, we present the first study of event-by-event jet drift via APE (Anisotropic Partonic Evolution), a new open-source 
Monte Carlo parton trajectory simulator, that incorporates a host of flexible model choices to survey the variability of the effect.
We demonstrate conclusively that even using the most conservative model choices
the imprint of flow on hard probes is robust and survives averaging over many events on the example of $\sqrt{s}=5.02$~TeV PbPb collisions at the Large Hadron Collider (LHC). We also show that well-motivated modifications to hadronization and medium modeling have the potential to further enhance the effect of jet drift, showing drastic modification to hard observables below hadronic $E$ of 10 GeV. Importantly, this new framework differs substantially from most other similar hard particle and jet simulators (e.g., HIJING \cite{Wang:1991hta}, HYDJET++ \cite{Lokhtin:2008xi}, JEWEL \cite{Zapp:2009ud}, JETSCAPE \cite{Putschke:2019yrg}, DREENA-A \cite{Zigic:2021rku}, HYBRID \cite{Casalderrey-Solana:2014bpa}) by tracking step-by-step average deflections to the trajectory of hard particles as they propagate through the evolving medium. In this work, we present the first rigorous, event-by-event simulation of flow effects from genuine power-suppressed corrections, which goes well beyond approaches which simply boost the leading-power (``eikonal'') scattering amplitudes to a moving frame \cite{Wang:1991hta, Armesto:2004pt, Casalderrey-Solana:2014bpa}.
%
\section{Methodology.}
%

Our realistic event-by-event medium is a re-implementation of the open-source DukeQCD hybrid transport model \cite{Moreland:2018gsh}, which consists of an initial condition from $T_R$ENTO \cite{Moreland:2014oya, Moreland:2017kdx}, 2+1D viscous hydrodynamics and Cooper-Frye particlization \cite{Song:2010aq} followed by the UrQMD hadronic afterburner \cite{Bass:1998ca}. Consistent with the boost-invariant medium, we treat the production of hard partons near mid-rapidity as boost invariant for a given $p_T$ \footnote{When discussing the power counting of the eikonal expansion, we refer to the parton energy $E = \sqrt{(p_T^{p})^2 + m^2} \cosh{y}$ rather than $p_T^p$ to avoid confusion with drift \textit{transverse to the jet axis}.  For light particle production at mid rapidity $y, m \approx 0$, the two are equivalent: $E \approx p_T^p$}. We employ the Maximum A Posteriori (MAP) parameters for the model from Ref.~\cite{Moreland:2018gsh}, and we verify that our simulations quantitatively reproduce the soft $v_2\{2\}$, $v_3\{2\}$, and $v_4\{2\}$ flow harmonics as well as identified hadron multiplicities, mean $p_T$, and $p_T$ fluctuations reported therein.

The seeds of jets in APE are modeled as single partons, for which we calculate the drift and energy loss.  This approach effectively neglects splittings other than medium-induced soft gluon emission, under the assumption that the hard scattering, in medium interactions, and fragmentation are factorizable~\cite{Vitev:2002pf}.  In many Monte Carlo simulations, a relatively small number $\sim\mathcal{O}(100)$ of QGP geometries are each seeded with many $\sim\mathcal{O}(10^6)$ 
hard-scattering events
for computational efficiency (see, e.g., \cite{Noronha-Hostler:2016eow, Jaki:Communication}). In contrast, we sample only 
$\sim\mathcal{O}(10^3)$ hard-scattering events
per event geometry, but we simulate a much larger number $\sim\mathcal{O}(10^5)$ of unique events. 
This lower oversampling factor retains the correlations within an event geometry as much as possible, while performing a much finer sampling of the event phase space.

Hard $2\rightarrow2$ processes from $\sqrt{s} = 5.02$ TeV Pythia \cite{Bierlich:2022pfr} pp collisions are sampled using an importance envelope of 
$(p_T^p/ 10 \: \text{GeV})^4$ for a parton with momentum $p_T^p$. Equivalently, this assigns a weight $w_i = (10 \: \text{GeV}/ p_T^p )^4$ 
to each trajectory, allowing us to maintain controlled statistics in the high-$p_T^{p}$ tail of the distribution. APE then embeds the unmodified partonic hard scattering products in the event geometry, distributed according to the binary collision density $n_{bc}(x,y) \propto s^2(x,y) \propto T^6(x,y)$ as in Refs.~\cite{Bernhard:2016tnd, Antiporda:2021hpk}, computing trajectories for a fixed grid of initial azimuthal angles for the scattering axis at each production point. The partons are initially antiparallel by construction, as we include only the tree-level scattering in this work.  This approximation does not affect single-inclusive distributions ($R_{AA}, v_2$) but does affect acoplanarities;  we thus  report the \textit{enhancement} due to the impact of drift, which may be modified with a fluctuating treatment of the broadening.

Jet seeds in the plasma are evolved along a path parameterized by pathlength $\ell$ using the following kinematic workflow:
\begin{enumerate}
    \setlength{\itemsep}{-4pt}
    \item Propagate the parton over a small timestep of length $\Delta t=0.1$ fm in the plasma center of mass rest frame.
    \item Compute the momentum transferred to the parton over the small pathlength $\Delta\ell = \beta \Delta t$ due to enabled effects, updating parton momentum accordingly.
    \item Repeat until the parton escapes the plasma perimeter at freezeout.
\end{enumerate}
We keep the quark masses,  noting that for the light flavors considered here $\beta \approx 1$. For every initial condition, we perform separate simulations with contributions of radiative energy loss, collisional energy loss, and jet drift effects switched on or off, resulting in fully-correlated distributions of hard particles for each combination of effects. This minimizes Monte Carlo noise in relative comparisons between them, allowing extraction of statistics-hungry measures like the elliptic flow \textit{enhancement} due to drift, $\Delta v_2^{exp}$ (Fig.~\ref{fig:deltav2}) in smaller sample sizes. Initial- and final- state properties of the hard partons are recorded before applying hadronization.
We consider no medium response effects at this time.

Neglecting the running of $\alpha_s$, we tune the coupling for the two energy loss cases (both including drift) via reduced $\chi^2$ fit to CMS $R_{AA}$ data at  30-50\% centrality and $p_T \in [8\; {\rm GeV}, 35 \; {\rm GeV}]$ \cite{CMS:2016xef}.  To two significant figures, we find $g = \sqrt{4 \pi \alpha_s} \approx 2.1$ with radiative energy loss alone and $g \approx 1.8$ with radiative and collisional energy loss, consistent with values from the literature~\cite{He:2011pd}. 
We have also explored the use of various forms of cold nuclear matter effects, but the observables reported here were found to be largely insensitive to their inclusion. Herein we choose to follow the perturbative treatment of Refs.~\cite{Vitev:2002pf, Gyulassy:2002yv, Qiu:2003vd} for the Cronin effect, isospin effects, and nuclear shadowing, but we neglect cold nuclear matter energy loss.

A perturbative calculation of parton drift and energy loss requires input for the mean free path $\lambda$ and Debye screening mass $\mu$. The latter reads
$   \mu \approx gT \sqrt{1 + {N_f}/{6}}$ where we use $N_f = 2$,
and by definition
${\lambda}^{-1} = \sigma_{q} \rho_{q} + \sigma_{g} \rho_{g}$,
with $\sigma_{q}$($\sigma_{g}$) the total cross section for the hard parton to interact with a quark (gluon) in the plasma and $\rho_{q}$($\rho_{q}$) the density of quarks (gluons) in the medium. Here we use the Gyulassy-Wang cross section modeled after static scattering centers in the medium~\cite{Gyulassy:1993hr}, and the  cross sections and partial quark and gluon densities are taken from~\cite{Sievert:2019cwq}, which approximates the medium as an ideal gas of gluons and quarks.
It should be noted that the identified parton number densities $\rho_q (\rho_g)$ are not provided by the hydrodynamic equation of state and are necessary to compute the various flavor-dependent channels of jet-medium interactions. We set $\lambda\to\infty$ when the medium 
cools below the hadronization temperature, which is chosen to be $T_{had}=155$ MeV. We note that this conservative assumption removes the potential effect of late-time interactions where collective flow, and thus drift, is strongest. 

Calculations of the energy loss of high-energy partons due to stimulated gluon emission are computationally demanding. This has led to a range of approaches to evaluate energy loss in realistic media. Many of the traditional works select an analytic form for the plasma \cite{Gyulassy:2000gk}; others create effective lengths and temperatures as realistic envelopes on an analytic model~\cite{Faraday:2023vbo}; while others simplify or parameterize the energy loss form \cite{Noronha-Hostler:2016eow}.
We elect to numerically compute the average single-emission Gyulassy-Levai-Vitev radiative energy loss at first order in opacity, including finite kinematic bounds for the momentum transfer to the medium and the radiated gluon momentum, using the Gyulassy-Wang form for the scattering potentials \cite{Gyulassy:2000gk}. The behavior of the radiative energy loss is qualitatively different at low $p_T$ for different choices of kinematic bounds, where sub-eikonal corrections become large. In a medium of effective length $L$, the mean energy loss rate is
\begin{align}   \label{e:ELossRate}
    & \frac{dE}{d\ell} = 
    - \frac{d}{dL} \Bigg(\frac{2 C_R \alpha_s}{\pi} \frac{L}{\lambda} E 
 \int_{{k}_{min}}^{{k}_{max}} \frac{dk}{k}  \; \int_0^{q_{max}} dq \, q \int_0^{2\pi} d\phi  \notag \\
    & \hspace{1cm} \times 
    \frac{\mu^2}{\pi (q^2 + \mu^2)^2} \frac{2 \mathbf{k}\cdot\mathbf{q} \, (\mathbf{k} - \mathbf{q})^2 L^2}{16 x^2 E^2 + (\mathbf{k} - \mathbf{q})^4 L^2} \Bigg)_{L=\ell} \, , 
\end{align}
with $k_{min} = \mu$, $k_{max} = \text{min}\{2Ex,2E\sqrt{x(1-x)}\}$, and $q_{max} = \sqrt{6ET}$ \cite{Gyulassy:2000er, Gyulassy:2000gk}. Here,  $C_R$ is the quadratic Casimir of the representation $R$ of SU(3) ($C_R = 3$ for gluons, $C_R = 4/3$ for quarks). We  take $\tau_{0} = 0.37$ fm for the thermalization time and do not treat the short-pathlength corrections as derived in Ref.~\cite{Kolbe:2015rvk}, which may be significant even for large collision systems. Thus, for massless partons, $L = \tau - \tau_0$, with $\tau$ the time relative to the collision in the plasma center of mass rest frame.

As described in Eq.~\eqref{e:ELossRate}, we evaluate the mean energy loss of a hard parton of energy $E$ in a QGP of fixed length and temperature, then compute the rate in $L$ by central differences. This yields the tabulated instantaneous energy loss rate $dE/d\ell $ we sample at runtime in each timestep of the partonic propagation. This method preserves the quadratic pathlength dependence associated with the Landau-Pomeranchuk-Migdal effect \cite{Migdal:1956tc, Landau:1953um} and the additional effects associated with the expanding medium, locally samples the properties of the QGP at each point of the partonic trajectory in a hydrodynamic geometry, and is computationally cheap at runtime.
\begin{figure}
    \centering
    \includegraphics[width=0.5\textwidth]{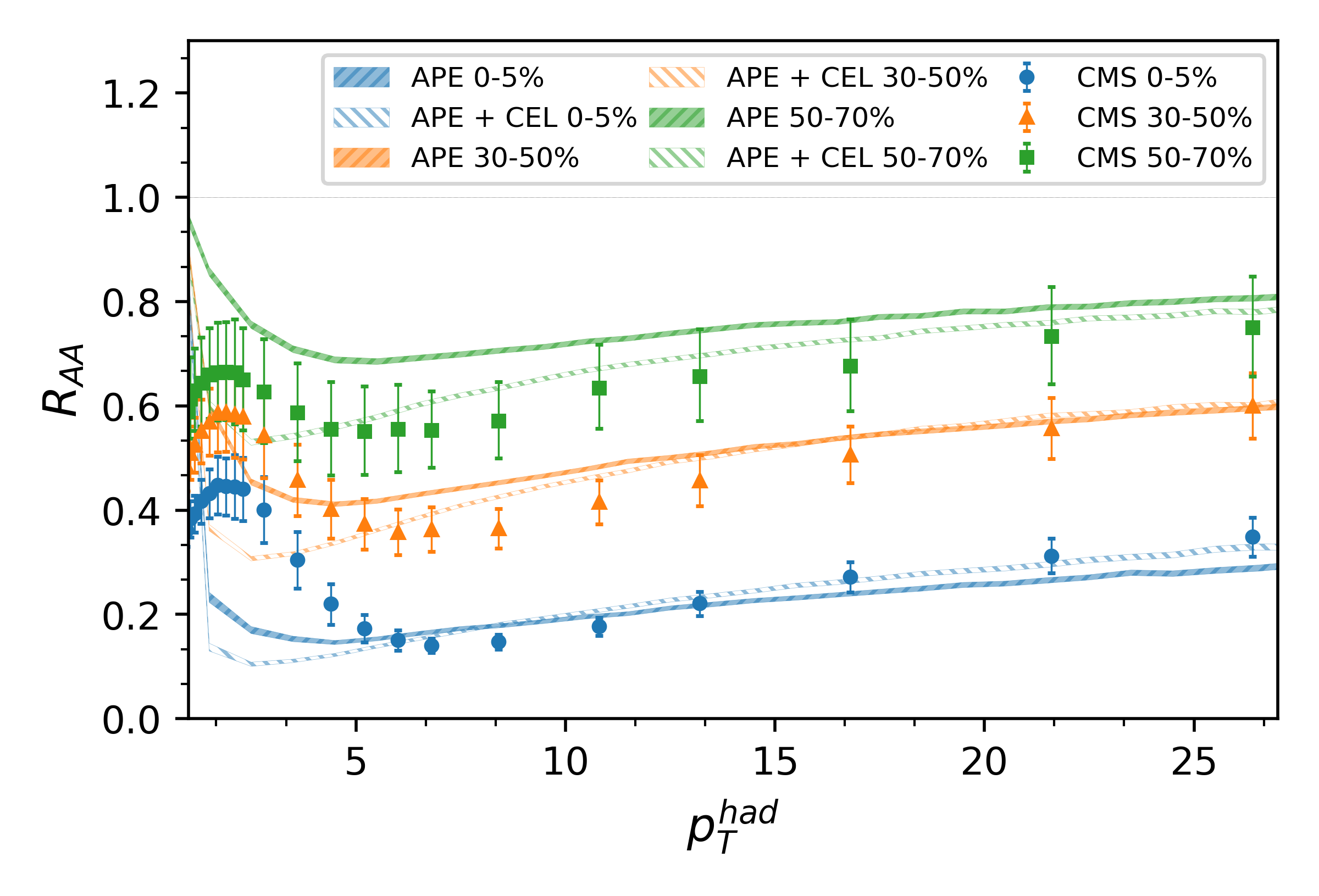}
    \caption{APE results for radiative energy loss alone and radiative energy loss + collisional energy loss (CEL) (bootstrap statistical error associated with the fluctuating event shown as hatched bands, with maximum value $\approx \pm 0.028$) plotted against CMS data for $R_{AA}(p_T^{had})$ at different centralities \cite{CMS:2016xef}.}
    \label{f:RAA}
\end{figure}
We implement an analytic calculation for the collisional energy loss \cite{Braaten:1991we} and can include it or neglect it in the computation, so as to estimate the effect on the phenomenology of drift:
\begin{align}   \label{e:EColl}
    & \frac{dE}{d\ell} = - \frac{C_R \, \mu^2}{2}  \ln\left( 2^{\frac{N_f}{2(6+N_f)}} 0.920 \frac{\sqrt{ET}}{m_g} \right),
\end{align}
where 
$m_g = \mu / \sqrt{3}$
is the thermal gluon mass. A finite-temperature perturbative calculation \cite{THOMA1991491} shows only minor corrections to Eq.~\eqref{e:EColl}. Likewise, the additional finite-$L$ corrections to Ref.~\cite{THOMA1991491} were also shown to be small \cite{Djordjevic:2006tw}.

We treat only the first odd moment of the flow-mediated drift computed in Ref.~\cite{Sadofyev:2021ohn}. The momentum transferred to the parton \textit{transverse to its trajectory} is
\begin{align}\label{e:q_drift_moment}
    \left\langle \vec{q}_{drift} \right\rangle & = \hat{e}_\perp \int d\ell \, \frac{3}{E} \, \frac{\mu^{2} }{\lambda} \, \ln\frac{E}{\mu} \: 
    \frac{u_\perp }{1-u_\parallel }.
\end{align}
Here, $\lambda$ is the mean free path, $\mu$ is the Debye screening  mass, and $\vec{u}$ is the collective flow velocity, with components $\parallel$ and $\perp$ to the parton momentum vector, and $\hat{e}_\perp$ is the corresponding unit vector.  Several recent works have extended the calculation of the impact of collective flow and gradients on medium-induced momentum broadening and radiation \cite{Barata:2022krd, Andres:2022ndd, Barata:2023qds, Kuzmin:2023hko, Kuzmin:2024smy}, however the moment we consider here has been shown to be exact to all orders in opacity \cite{Andres:2022ndd}.

The distribution of leading hadrons can be calculated from the distribution of leading partons by convolution with fragmentation functions (FFs) $D_{p}^{h}(z)$ \cite{Markert:2008jc}
\begin{align} \label{e:sigma_diff}
    \frac{d\sigma_{NN}^{h}}{dydp_{T}^{h}} & = \int dz \frac{1}{z} \, D^{h}_{p}(z) \, \frac{d\sigma_{NN}^{p}}{dydp_{T}^{p}}.
\end{align}
In order to evaluate this expression for a discrete distribution of quarks and gluons, we perform an event-by-event fragmentation  such that for a given parton with fixed flavor and $p_T^{p}$, we sample from a probability distribution function for hadronic momentum fraction $z = p_T^{h}/p_T^{p}$ proportional to the inclusive fragmentation function, 
$P_{p_{T}^{p}}(z) \propto D_{p}^{h}(z)$.
We compute the distribution of inclusive charged hadrons, applying the JAM Collaboration's ``JAM20-SIDIS" FF fits for charged hadrons \cite{Moffat:2021dji} accessed through LHAPDF \cite{Buckley:2014ana}.

This treatment of the system as a whole takes the most conservative choices for each phase of the computation. The results represent a realistic lower-bound on the effect size of flow mediated drift. This energy regime, and especially the flow-mediated drift effect, is quite sensitive to model choices at every step. To study the impact of various other viable scenarios, we also implement a sharper, more anisotropic optical Glauber initial condition for the medium and formulate a simple treatment of hybrid coalescence effects.
For the initial entropy density of the sharp optical Glauber medium model, we take the harmonic mean of two Woods-Saxon nuclear density profiles $T_{A,B}$ \cite{Moreland:2014oya, Moreland:2017kdx},  $ {dS}/{dy} \propto {2 T_A T_B}{/(T_A + T_B)}$.
The higher anisotropy and sharper gradients of these initial conditions produce larger deflections due to flow. The medium evolution is treated identically.
Following \cite{Greco:2003xt}, we presume that pions dominate the charged hadron distribution, thus we take the distribution of hadrons formed from the hybrid coalescence of a hard parton and a thermal parton to be
\begin{align}
    \frac{dN_{h}}{d p_{h}d\phi_{h}} &= g_{\pi}\frac{6\pi^2}{V\Delta_p^3}
    \int d p_1 d\phi_1 \, d p_{2} d\phi_{2}  \notag \\
    &\hspace{1cm} \times \frac{dN_{hard}}{d p_1 d\phi_1} \: \frac{dN_{thermal}}{d p_{2}d\phi_{2}} \:
    \delta^3 (\vec{p}_h - \vec{p}_1 - \vec{p}_{2} ) \notag \\
    &\hspace{1cm} \times \Theta(\Delta_p - |p_{T, 1} - p_{T, 2}|/2) , 
\end{align}
where we have taken the same model form for the Wigner distribution of the pion as \cite{Greco:2003xt} (with the momentum width of the pion $\Delta_p = 0.24$ GeV, $V$ the volume of the QGP, and $g_{\pi}$ the formation probability). Our thermal parton distribution is a simple Boltzmann distribution in the center of mass rest frame of the fluid with the appropriate boson or fermion statistics for the hard parton's color neutralizing counterpart, $dN_{thermal}/dp_1d\phi_1 \propto 1 / (e^{E/T_{had}} \pm 1)$

%
\section{Results.}
%

A simultaneous description of the suppression $R_{AA} (p_T)$ of jets and their elliptic modulation, $v_2(p_T)$, has been one of the benchmark problems of jet physics in heavy ion collisions \cite{Noronha-Hostler:2016eow, Molnar:2013eqa, Zhao:2021vmu, Zhang:2013oca, Kopeliovich:2012sc, Andres:2019eus, PhysRevD.108.074013}. Various microscopic models and calculations can simultaneously capture both observables in the high-energy regime, $p_T > 10$ GeV, but perturbative calculations have not successfully described the intermediate and low energy regimes, $p_T < 10$ GeV, where elliptic flow is generally underpredicted \cite{Molnar:2013eqa, Zhang:2013oca}. 

Jet drift is a perturbative correction to the transverse momentum distribution that naturally supplies some of the missing flow.  While this sub-eikonal effect does not directly modify the energy loss, 
drift may indirectly affect the suppression by deflecting the paths toward regions of lower density. This deflection, however, directly modifies the azimuthal distribution of particles. The coupling of drift to the medium correlates leading hadrons to the event plane by dragging hard particles towards ``attractors'' at the minor axes of the elliptic medium geometry, as schematically shown in Fig~\ref{fig:sample_trajectory}. 

\begin{figure}
\begin{centering}
\includegraphics[width=0.5\textwidth]{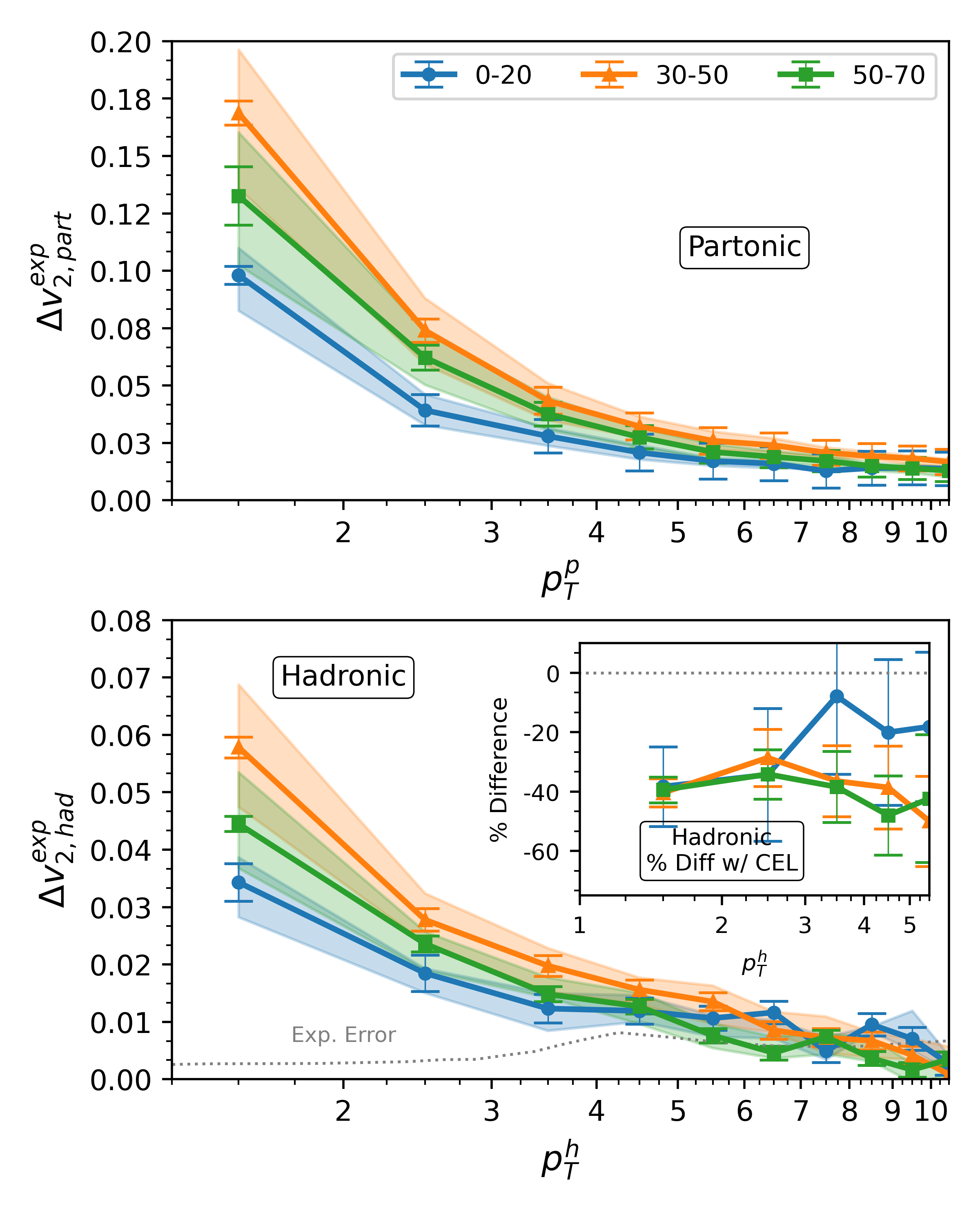}
    \caption{
    $\Delta v_2^{exp}$ from flow-mediated jet drift as a function of leading-partonic $p_T^{p}$ (upper panel) and leading-hadronic $p_T^{h}$ (lower panels) in several centrality bins, including the percent difference of hadronic $\Delta v_2^{exp}$ with collisional and radiative energy loss to only radiative (inset). Theoretical error is quantified by varied strength of drift from 75-125\%, and bootstrap statistical error associated with the fluctuating event is shown in fences. Experimental error threshold is plotted for 30-40\% centrality charged pions from CMS \cite{CMS:2017xgk}, which is everywhere $\leq 1\%$. The $p_T$ bin width is fixed to be 1 GeV.}
    \label{fig:deltav2}
\end{centering}
\end{figure}

We report the suppression $R_{AA}$ as computed from Eq.~\eqref{e:RAA} in Fig~\ref{f:RAA} and the enhancement of elliptic flow computed from Eq.~\eqref{e:vndefn} as the difference in elliptic flow with and without drift $\Delta v_2^{exp}$ in Fig~\ref{fig:deltav2}.  We take the denominator of Eq.~\eqref{e:RAA} as the number distribution of partons seeded from Pythia and compute the complex vector $\vec{v}_2^{hard}$ via the Q-cumulant method \cite{Bilandzic:2012wva} as a discrete sum over all trajectories $i$ producing final state particles in the given $p_T$ bin.
\begin{align}
    \vec{v}_n^{hard}(p_T) & = \left\langle \frac{ \sum_{i} w_i 
   e^{i n \phi_i }}{\sum_{i} w_i}\right\rangle,
\end{align}
where $w_i$ is the weight associated with parton $i$ and $\phi_i$ the final azimuthal angle of the trajectory.
Finally, the angle brackets represent an average over events. $\vec{v}_2^{soft}$ is computed as for $\vec{v}_2^{hard}$ for thermal particles $p_T \in [0.2\; {\rm GeV}, 5 \; {\rm GeV}]$ from UrQMD. $v_n^{exp}(p_T)$ is the scalar product of the soft and hard flow vectors
\begin{align} \label{e:vndefn}
    {v}_n^{exp}(p_T) & =  \frac{\left\langle \vec{v}_n^{hard} \cdot \vec{v}_n^{soft} \right\rangle}{\sqrt{\left\langle \left|\vec{v}_n^{soft}\right|^2\right\rangle}}.
\end{align}

Intriguingly, we obtain a nuclear modification factor $R_{AA} (p_T)$ in Fig.~\ref{f:RAA} which is unmodified due to the effects of drift.  This suggests that any  systematic bias of trajectories toward lower density is not significant for the nuclear modification factor.  
We do, however, see the anticipated enhancement of the elliptic flow shown in Fig.~\ref{fig:deltav2}, which turns out to be sizeable: an enhancement of several times the experimental uncertainty (e.g. $< \pm 0.005$ for charged pion $v_2$ at 3.00 - 3.25 GeV in the 30-40\% bin \cite{ALICE:2018yph}). Interestingly, the enhancement from drift qualitatively matches the excess in elliptic flow seen in experiment below 10 GeV \cite{CMS:2017xgk}.  While we make no claim that jet drift is \textit{the} solution to the $R_{AA} \otimes v_2$ puzzle, it must nevertheless be an important \textit{part} of the solution in the $p_T < 10$ GeV regime.
We also note that the jet drift effect is strongest in semi-peripheral collisions (30-50\% centrality), which balance a larger event eccentricity with the decreasing event temperatures in smaller-sized QGP. This demonstrates the interplay between energy loss and drift, whereby energy loss enhances the strength of drift, but strong energy loss extinguishes partons that have experienced strong drift. Likewise, the addition of collisional energy loss (and the refitting of the coupling) reduces the strength of the $v_2^{hard}$ enhancement by about 50\%, as shown in the ratio inset of Fig.~\ref{fig:deltav2}, as a lower overall coupling is necessary to fit high-$p_T$ suppression, and the relative strength of drift is reduced. The addition of more sources of energy loss (including from cold nuclear matter effects) generally reduces the relative strength of drift, however a more realistic fluctuating treatment of collisional energy loss may not display this same behavior.

\begin{figure}
\begin{centering}
\includegraphics[width=0.5\textwidth]{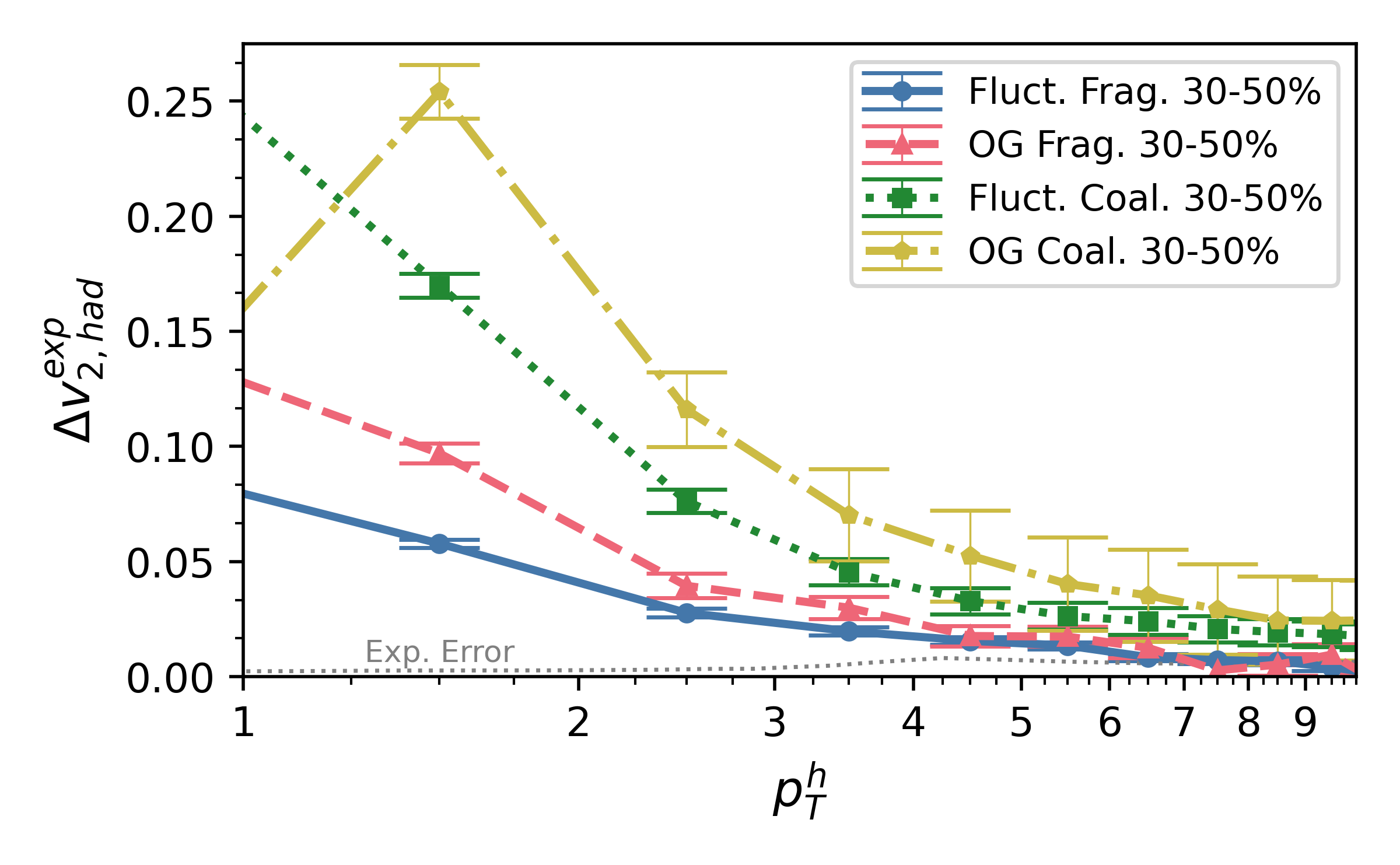}
\caption{Comparison of $\Delta v_2^{exp}$ in APE 30-50\% centrality collisions for combinations of optical Glauber vs fluctuating initial conditions and coalescence vs fragmentation hadronization mechanisms. Bootstrap statistical error associated with the fluctuating event is shown in fences. Experimental error threshold is plotted for 30-40\% centrality charged pions from CMS \cite{CMS:2017xgk}, which is everywhere $\leq 1\%$. The $p_T$ bin width is fixed to be 1 GeV.
\label{fig:optimism}
}
\end{centering}
\end{figure}

The strength of the anisotopic flow modification ($\Delta v_2^{exp}$) is significantly enhanced by the inclusion of coalescence or the use of a sharper initial condition for the medium evolution. As shown in Fig.~\ref{fig:optimism}, coalescence upshifts the $p_T$ of strongly drifted partons, expanding the region of interest and dramatically increasing the magnitude of the enhancement. This implementation takes an isotropic distribution of thermal partons, and thus does not include direct convolution with the azimuthal anisotropy of the medium, which is expected to have additional complex interplay with drift. The combination of both changes represents an optimistic, rational estimate, but is by no means an upper bound on the effect. 

\begin{figure}
\begin{centering}
\includegraphics[width=0.5\textwidth]{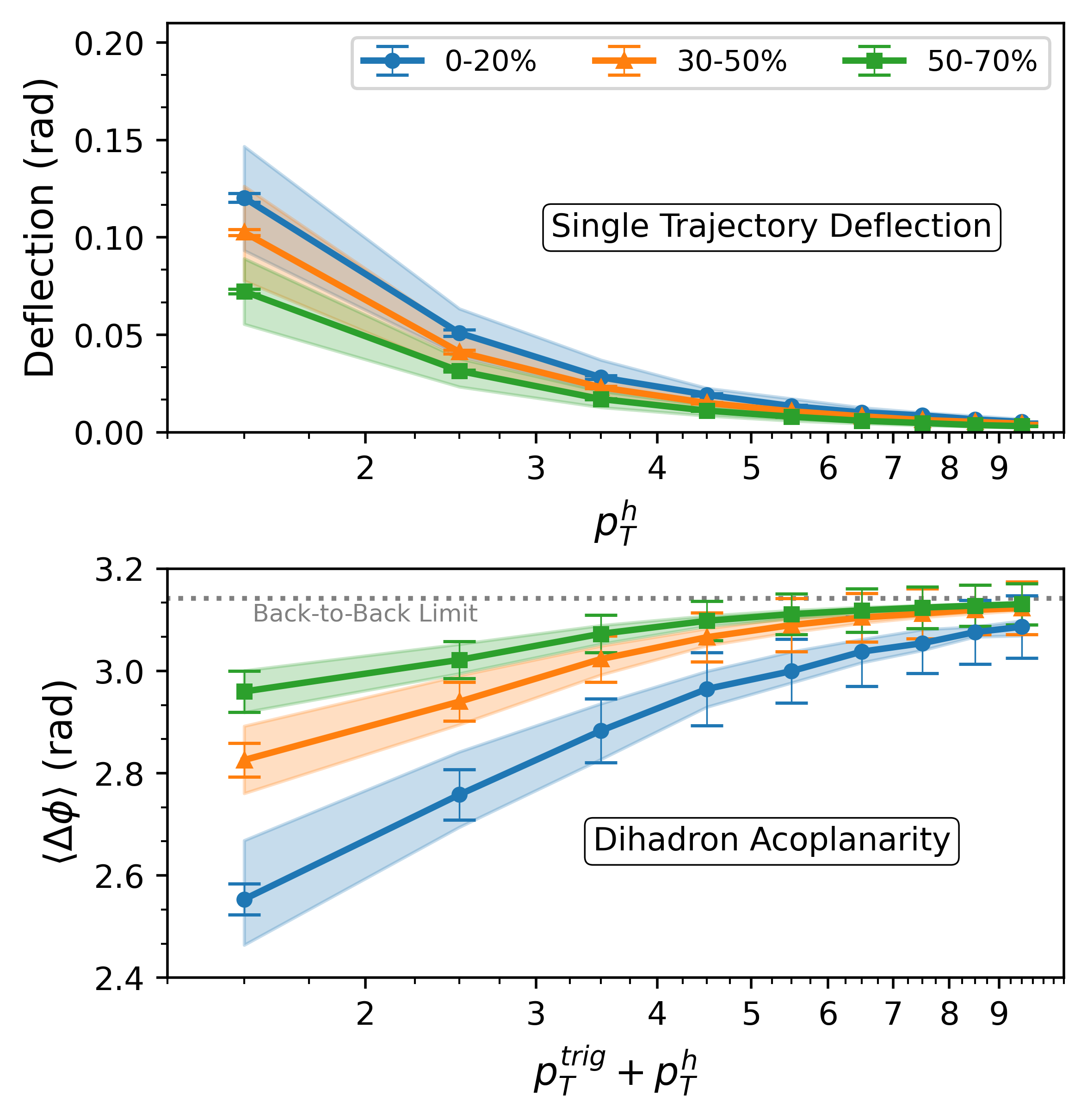}
\caption{
Top: Mean angle of deflection due to drift as a function of $p_T^{h}$ for charged hadrons. 
Bottom: Mean dihadron acoplanarity (deviation from $\pi$), including \textit{only} anisotropic broadening and radiative energy loss as a function of $p_T^{trig} + p_T^{h}$. The $p_T$ bin width is fixed to be 1 GeV. Bootstrap statistical error associated with the fluctuating event is shown in fences, with maximum deflection error value $\approx \pm 0.002$.
\label{fig:acoplanarities}
}
\end{centering}
\end{figure}

We further study acoplanarities in APE by embedding the pairs of partons from each hard process on exactly antiparallel trajectories at the same production point in a given event geometry.  The resulting deflections of single partons, defined as the difference in initial and final trajectory angle, and acoplanarities, defined as the smallest angle between parton pairs from the same hard process in the final state, are plotted in Fig.~\ref{fig:acoplanarities}.

The correlated deflections of particles from drifting in the same event geometry results in an  enhancement of the acoplanarity (a deviation from back-to-back) (Fig.~\ref{fig:acoplanarities}) between the hadrons.  The additional contribution from drift can be easily understood from Fig.~\ref{fig:sample_trajectory}, in which widely-separated parton trajectories are deflected toward the same ``attractor" in the event geometry.  While $v_2^{hard}$ measures deflections \textit{correlated with the event plane}, the azimuthally averaged acoplanarity provides a complementary measure of the \textit{magnitude} of the deflections, independent of the event plane, although the strength of the enhancement varies azimuthally relative to the event plane.

Like the elliptic flow $v_2$, the mean deflection angles and acoplanarities show sizeable enhancements due to drift at low $p_T$, e.g. greater than 5 degree acoplanarity below 6 GeV.  Importantly, however, the centrality dependence has changed between Fig.~\ref{fig:deltav2} and Fig.~\ref{fig:acoplanarities}.  The elliptic flow enhancement $\Delta v_2^{exp}$ is largest for the intermediate $30-50\%$ centrality bin, while the deflections and acoplanarities are largest in the $0-20\%$ central bin.  This is a reflection of the fact that acoplanarities are sensitive to the \textit{magnitude} of drift, which is largest in central collisions but more weakly correlated with the event plane.  The elliptic flow, on the other hand, reflects not only the temperature but also the ellipticity, which is largest in peripheral collisions.

%
\section{Conclusions.\label{sec:conclusions}}
%

Flow-mediated drift, as studied here, belongs both to a family of sub-eikonal effects that extend the theory of jet quenching toward lower $p_T$, and to a family of antisymmetric effects that can generate azimuthal anisotropies of hard probes. This work shows that anisotropic effects modify hard observables in ways that survive event averaging, and sub-eikonal effects can systematically extend perturbative calculations to  intermediate- and low-$p_T$. 
We have further demonstrated that inclusive measures like $v_2$ and correlations like acoplanarities exhibit different systematics in their response to drift.
Jet drift and similar effects are sensitive to properties of the medium 
that neither isotropic measures of hard probes nor soft-sector anisotropies can capture, including local flow structure, flowing hot spots, and other internal structural differences. 
These novel properties of jet drift  provide an additional dimension of sensitivity with which to expand the prospects of jet QGP tomography championed by the field.

Throughout this study, the main results presented a conservative estimate of the effect of jet drift. spite the unfavorable assumptions as described in our methodology section, we show here that jet drift nonetheless leaves measurable imprints of the flow on low-$p_T^{h}$ particles, systematically enhancing elliptic flow and increasing dijet acoplanarity. The reasonable inclusion of coalescence and use of a sharper medium model significantly enhances the effect  and  serves to emphasize how sensitive the results of an intermediate $p_T$ simulation are to model choices, especially in regards to drift.
Consequently, this study constitutes the first unambiguous demonstration of the importance of full velocity tomography in event-by-event heavy-ion collisions and the first comprehensive phenomenological investigation of the flow modification of hard probes.

In the future, we plan to study the implications of drift for  intra-jet particle and energy distributions \cite{Ke:2024emw}. 
Previous work has suggested that gradients and flow may impact the distribution of radiation in the jet cone \cite{Armesto:2004pt, Sadofyev:2021ohn, Barata:2023zqg}, however we expect jet substructure to be even more sensitive to flow than the inclusive observables considered here due to the \textit{combination} of this asymmetric gluon radiation and deflections due to drift.

\section{Acknowledgments.\label{sec:acknowledgments}}
%
%
\begin{acknowledgments}
    This material is based upon work supported by the U.S. Department of Energy, Office of Science, Office of Nuclear Physics under Award Number DE-SC0024560 (JB, HR, MS, IV) and the Los Alamos National Laboratory (Contract No. 89233218CNA000001) and its LDRD Program under Award Number 20240131ER (IV). This work utilized resources from the New Mexico State University High Performance Computing Group, which is directly supported by the National Science Foundation (OAC-2019000), the Student Technology Advisory Committee, and New Mexico State University and benefits from inclusion in various grants (DoD ARO-W911NF1810454; NSF EPSCoR OIA-1757207; Partnership for the Advancement of Cancer Research, supported in part by NCI grants U54 CA132383 (NMSU)).  This research utilized services provided by the OSG Consortium \cite{osg07, osg09, https://doi.org/10.21231/906p-4d78, https://doi.org/10.21231/0kvz-ve57}, which is supported by the National Science Foundation awards \#2030508 and \#1836650.
    The open-source APE software suite and associated resources are publicly available on \href{https://github.com/Jopacabra/ape}{GitHub}. JB and IV acknowledge INT support. 
\end{acknowledgments}
\bibliographystyle{bibstyle}
\bibliography{Refs}

\begin{thebibliography}{10}
\ifx\href\asklfhas\newcommand{\href}[2]{#2}\fi
\ifx\arxivref\asklfhas\newcommand{\arxivref}[2]{\href{http://arxiv.org/abs/#1}{#2}}\fi
\ifx\doiref\asklfhas\newcommand{\doiref}[2]{\href{http://dx.doi.org/#1}{#2}}\fi
\parskip 0pt
\normalsize

\bibitem{Connors:2017ptx}
M.~Connors, C.~Nattrass, R.~Reed \& S.~Salur,
\textit{``{Jet measurements in heavy ion physics}''},
\doiref{10.1103/RevModPhys.90.025005}{Rev.~Mod.~Phys. \textbf{90}, 025005
  (2018)},
\normalsize{\texttt{\arxivref{1705.01974}{arXiv:1705.01974}}}.

\bibitem{Vitev:2002pf}
I.~Vitev \& M.~Gyulassy,
\textit{``{High $p_{T}$ tomography of $d$ + Au and Au+Au at SPS, RHIC, and
  LHC}''},
\doiref{10.1103/PhysRevLett.89.252301}{Phys.~Rev.~Lett. \textbf{89}, 252301
  (2002)},
\normalsize{\texttt{\arxivref{hep-ph/0209161}{hep-ph/0209161}}}.

\bibitem{Noronha-Hostler:2016eow}
J.~Noronha-Hostler, B.~Betz, J.~Noronha \& M.~Gyulassy,
\textit{``{Event-by-event hydrodynamics $+$ jet energy loss: A solution to the
  $R_{AA} \otimes v_2$ puzzle}''},
\doiref{10.1103/PhysRevLett.116.252301}{Phys.~Rev.~Lett. \textbf{116}, 252301
  (2016)},
\normalsize{\texttt{\arxivref{1602.03788}{arXiv:1602.03788}}}.

\bibitem{Molnar:2013eqa}
D.~Molnar \& D.~Sun,
\textit{``{High-pT suppression and elliptic flow from radiative energy loss
  with realistic bulk medium expansion}''},
\normalsize{\texttt{\arxivref{1305.1046}{arXiv:1305.1046}}}.

\bibitem{Zhao:2021vmu}
W.~Zhao, W.~Ke, W.~Chen, T.~Luo \& X.-N. Wang,
\textit{``{From Hydrodynamics to Jet Quenching, Coalescence, and Hadron
  Cascade: A Coupled Approach to Solving the RAA\ensuremath{\otimes}v2
  Puzzle}''},
\doiref{10.1103/PhysRevLett.128.022302}{Phys.~Rev.~Lett. \textbf{128}, 022302
  (2022)},
\normalsize{\texttt{\arxivref{2103.14657}{arXiv:2103.14657}}}.

\bibitem{Zhang:2013oca}
X.~Zhang \& J.~Liao,
\textit{``{Jet Quenching and Its Azimuthal Anisotropy in AA and possibly High
  Multiplicity pA and dA Collisions}''},
\normalsize{\texttt{\arxivref{1311.5463}{arXiv:1311.5463}}}.

\bibitem{Kopeliovich:2012sc}
B.~Z. Kopeliovich, J.~Nemchik, I.~K. Potashnikova \& I.~Schmidt,
\textit{``{Quenching of high-pT hadrons: Energy Loss vs Color Transparency}''},
\doiref{10.1103/PhysRevC.86.054904}{Phys.~Rev.~C \textbf{86}, 054904 (2012)},
\normalsize{\texttt{\arxivref{1208.4951}{arXiv:1208.4951}}}.

\bibitem{Andres:2019eus}
C.~Andres, N.~Armesto, H.~Niemi, R.~Paatelainen \& C.~A. Salgado,
\textit{``{Jet quenching as a probe of the initial stages in heavy-ion
  collisions}''},
\doiref{10.1016/j.physletb.2020.135318}{Phys.~Lett.~B \textbf{803}, 135318
  (2020)},
\normalsize{\texttt{\arxivref{1902.03231}{arXiv:1902.03231}}}.

\bibitem{CMS:2024krd}
CMS Collaboration, A.~Hayrapetyan et~al.,
\textit{``{Overview of high-density QCD studies with the CMS experiment at the
  LHC}''},
\normalsize{\texttt{\arxivref{2405.10785}{arXiv:2405.10785}}}.

\bibitem{ALICE:2023qve}
ALICE Collaboration, S.~Acharya et~al.,
\textit{``{Observation of Medium-Induced Yield Enhancement and Acoplanarity
  Broadening of Low-pT Jets from Measurements in pp and Central Pb-Pb
  Collisions at sNN=5.02\,\,TeV}''},
\doiref{10.1103/PhysRevLett.133.022301}{Phys.~Rev.~Lett. \textbf{133}, 022301
  (2024)},
\normalsize{\texttt{\arxivref{2308.16131}{arXiv:2308.16131}}}.

\bibitem{Sadofyev:2021ohn}
A.~V. Sadofyev, M.~D. Sievert \& I.~Vitev,
\textit{``{Ab Initio Coupling of Jets to Collective Flow in the Opacity
  Expansion Approach}''},
\normalsize{\texttt{\arxivref{2104.09513}{arXiv:2104.09513}}}.

\bibitem{Antiporda:2021hpk}
L.~Antiporda, J.~Bahder, H.~Rahman \& M.~D. Sievert,
\textit{``{Jet Drift and Collective Flow in Heavy-Ion Collisions}''},
\normalsize{\texttt{\arxivref{2110.03590}{arXiv:2110.03590}}}.

\bibitem{Wang:1991hta}
X.-N. Wang \& M.~Gyulassy,
\textit{``{HIJING: A Monte Carlo model for multiple jet production in p p, p A
  and A A collisions}''},
\doiref{10.1103/PhysRevD.44.3501}{Phys.~Rev.~D \textbf{44}, 3501 (1991)}.

\bibitem{Lokhtin:2008xi}
I.~P. Lokhtin, L.~V. Malinina, S.~V. Petrushanko, A.~M. Snigirev, I.~Arsene \&
  K.~Tywoniuk,
\textit{``{Heavy ion event generator HYDJET++ (HYDrodynamics plus JETs)}''},
\doiref{10.1016/j.cpc.2008.11.015}{Comput.~Phys.~Commun. \textbf{180}, 779
  (2009)},
\normalsize{\texttt{\arxivref{0809.2708}{arXiv:0809.2708}}}.

\bibitem{Zapp:2009ud}
K.~Zapp, J.~Stachel \& U.~A. Wiedemann,
\textit{``{JEWEL - a Monte Carlo Model for Jet Quenching}''},
\doiref{10.22323/1.080.0022}{PoS \textbf{High-pT physics09}, 022 (2009)},
\normalsize{\texttt{\arxivref{0904.4885}{arXiv:0904.4885}}}.

\bibitem{Putschke:2019yrg}
J.~H. Putschke et~al.,
\textit{``{The JETSCAPE framework}''},
\normalsize{\texttt{\arxivref{1903.07706}{arXiv:1903.07706}}}.

\bibitem{Zigic:2021rku}
D.~Zigic, I.~Salom, J.~Auvinen, P.~Huovinen \& M.~Djordjevic,
\textit{``{DREENA-A framework as a QGP tomography tool}''},
\doiref{10.3389/fphy.2022.957019}{Front.~in~Phys. \textbf{10}, 957019 (2022)},
\normalsize{\texttt{\arxivref{2110.01544}{arXiv:2110.01544}}}.

\bibitem{Casalderrey-Solana:2014bpa}
J.~Casalderrey-Solana, D.~C. Gulhan, J.~G. Milhano, D.~Pablos \& K.~Rajagopal,
\textit{``{A Hybrid Strong/Weak Coupling Approach to Jet Quenching}''},
\doiref{10.1007/JHEP09(2015)175}{JHEP \textbf{1410}, 019 (2014)},
\normalsize{\texttt{\arxivref{1405.3864}{arXiv:1405.3864}}},
[Erratum: JHEP 09, 175 (2015)].

\bibitem{Armesto:2004pt}
N.~Armesto, C.~A. Salgado \& U.~A. Wiedemann,
\textit{``{Measuring the collective flow with jets}''},
\doiref{10.1103/PhysRevLett.93.242301}{Phys.~Rev.~Lett. \textbf{93}, 242301
  (2004)},
\normalsize{\texttt{\arxivref{hep-ph/0405301}{hep-ph/0405301}}}.

\bibitem{Moreland:2018gsh}
J.~S. Moreland, J.~E. Bernhard \& S.~A. Bass,
\textit{``{Bayesian calibration of a hybrid nuclear collision model using p-Pb
  and Pb-Pb data at energies available at the CERN Large Hadron Collider}''},
\doiref{10.1103/PhysRevC.101.024911}{Phys.~Rev.~C \textbf{101}, 024911 (2020)},
\normalsize{\texttt{\arxivref{1808.02106}{arXiv:1808.02106}}}.

\bibitem{Moreland:2014oya}
J.~S. Moreland, J.~E. Bernhard \& S.~A. Bass,
\textit{``{Alternative ansatz to wounded nucleon and binary collision scaling
  in high-energy nuclear collisions}''},
\doiref{10.1103/PhysRevC.92.011901}{Phys.~Rev. \textbf{C92}, 011901 (2015)},
\normalsize{\texttt{\arxivref{1412.4708}{arXiv:1412.4708}}}.

\bibitem{Moreland:2017kdx}
J.~S. Moreland, J.~E. Bernhard, W.~Ke \& S.~A. Bass,
\textit{``{Flow in small and large quark-gluon plasma droplets: the role of
  nucleon substructure}''},
\doiref{10.1016/j.nuclphysa.2017.05.054}{Nucl.~Phys.~A \textbf{967}, 361
  (2017)},
\normalsize{\texttt{\arxivref{1704.04486}{arXiv:1704.04486}}}.

\bibitem{Song:2010aq}
H.~Song, S.~A. Bass \& U.~Heinz,
\textit{``{Viscous QCD matter in a hybrid hydrodynamic+Boltzmann approach}''},
\doiref{10.1103/PhysRevC.83.024912}{Phys.~Rev.~C \textbf{83}, 024912 (2011)},
\normalsize{\texttt{\arxivref{1012.0555}{arXiv:1012.0555}}}.

\bibitem{Bass:1998ca}
S.~A. Bass et~al.,
\textit{``{Microscopic models for ultrarelativistic heavy ion collisions}''},
\doiref{10.1016/S0146-6410(98)00058-1}{Prog.~Part.~Nucl.~Phys. \textbf{41}, 255
  (1998)},
\normalsize{\texttt{\arxivref{nucl-th/9803035}{nucl-th/9803035}}}.

\bibitem{Note1}
When discussing the power counting of the eikonal expansion, we refer to the
  parton energy $E = \protect \sqrt {(p_T^{p})^2 + m^2} \cosh {y}$ rather than
  $p_T^p$ to avoid confusion with drift \protect \textit {transverse to the jet
  axis}. For light particle production at mid rapidity $y, m \approx 0$, the
  two are equivalent: $E \approx p_T^p$.

\bibitem{Jaki:Communication}
J.~Noronha-Hostler.

\bibitem{Bierlich:2022pfr}
C.~Bierlich et~al.,
\textit{``{A comprehensive guide to the physics and usage of PYTHIA 8.3}''},
\normalsize{\texttt{\arxivref{2203.11601}{arXiv:2203.11601}}}.

\bibitem{Bernhard:2016tnd}
J.~E. Bernhard, J.~S. Moreland, S.~A. Bass, J.~Liu \& U.~Heinz,
\textit{``{Applying Bayesian parameter estimation to relativistic heavy-ion
  collisions: simultaneous characterization of the initial state and
  quark-gluon plasma medium}''},
\doiref{10.1103/PhysRevC.94.024907}{Phys.~Rev. \textbf{C94}, 024907 (2016)},
\normalsize{\texttt{\arxivref{1605.03954}{arXiv:1605.03954}}}.

\bibitem{CMS:2016xef}
CMS Collaboration, V.~Khachatryan et~al.,
\textit{``{Charged-particle nuclear modification factors in PbPb and pPb
  collisions at $ \sqrt{s_{\mathrm{N}\;\mathrm{N}}}=5.02 $ TeV}''},
\doiref{10.1007/JHEP04(2017)039}{JHEP \textbf{1704}, 039 (2017)},
\normalsize{\texttt{\arxivref{1611.01664}{arXiv:1611.01664}}}.

\bibitem{He:2011pd}
Y.~He, I.~Vitev \& B.-W. Zhang,
\textit{``{${\cal O}(\alpha_s^3)$ Analysis of Inclusive Jet and di-Jet
  Production in Heavy Ion Reactions at the Large Hadron Collider}''},
\doiref{10.1016/j.physletb.2012.05.054}{Phys.~Lett.~B \textbf{713}, 224
  (2012)},
\normalsize{\texttt{\arxivref{1105.2566}{arXiv:1105.2566}}}.

\bibitem{Gyulassy:2002yv}
M.~Gyulassy, P.~Levai \& I.~Vitev,
\textit{``{Reaction operator approach to multiple elastic scatterings}''},
\doiref{10.1103/PhysRevD.66.014005}{Phys.~Rev.~D \textbf{66}, 014005 (2002)},
\normalsize{\texttt{\arxivref{nucl-th/0201078}{nucl-th/0201078}}}.

\bibitem{Qiu:2003vd}
J.-w. Qiu \& I.~Vitev,
\textit{``{Resummed QCD power corrections to nuclear shadowing}''},
\doiref{10.1103/PhysRevLett.93.262301}{Phys.~Rev.~Lett. \textbf{93}, 262301
  (2004)},
\normalsize{\texttt{\arxivref{hep-ph/0309094}{hep-ph/0309094}}}.

\bibitem{Gyulassy:1993hr}
M.~Gyulassy \& X.-n. Wang,
\textit{``{Multiple collisions and induced gluon Bremsstrahlung in QCD}''},
\doiref{10.1016/0550-3213(94)90079-5}{Nucl.~Phys. \textbf{B420}, 583 (1994)},
\normalsize{\texttt{\arxivref{nucl-th/9306003}{nucl-th/9306003}}}.

\bibitem{Sievert:2019cwq}
M.~D. Sievert, I.~Vitev \& B.~Yoon,
\textit{``{A complete set of in-medium splitting functions to any order in
  opacity}''},
\doiref{10.1016/j.physletb.2019.06.019}{Phys.~Lett.~B \textbf{795}, 502
  (2019)},
\normalsize{\texttt{\arxivref{1903.06170}{arXiv:1903.06170}}}.

\bibitem{Gyulassy:2000gk}
M.~Gyulassy, I.~Vitev \& X.~N. Wang,
\textit{``{High p(T) azimuthal asymmetry in noncentral A+A at RHIC}''},
\doiref{10.1103/PhysRevLett.86.2537}{Phys.~Rev.~Lett. \textbf{86}, 2537
  (2001)},
\normalsize{\texttt{\arxivref{nucl-th/0012092}{nucl-th/0012092}}}.

\bibitem{Faraday:2023vbo}
C.~Faraday \& W.~Horowitz,
\textit{``{Energy Loss in Small Collision Systems}''},
\doiref{10.22323/1.438.0131}{PoS \textbf{HardProbes2023}, 131 (2024)},
\normalsize{\texttt{\arxivref{2307.08355}{arXiv:2307.08355}}}.

\bibitem{Gyulassy:2000er}
M.~Gyulassy, P.~Levai \& I.~Vitev,
\textit{``{Reaction operator approach to nonAbelian energy loss}''},
\doiref{10.1016/S0550-3213(00)00652-0}{Nucl.Phys. \textbf{B594}, 371 (2001)},
\normalsize{\texttt{\arxivref{nucl-th/0006010}{nucl-th/0006010}}}.

\bibitem{Kolbe:2015rvk}
I.~Kolbe \& W.~A. Horowitz,
\textit{``{Short path length corrections to Djordjevic-Gyulassy-Levai-Vitev
  energy loss}''},
\doiref{10.1103/PhysRevC.100.024913}{Phys.~Rev.~C \textbf{100}, 024913 (2019)},
\normalsize{\texttt{\arxivref{1511.09313}{arXiv:1511.09313}}}.

\bibitem{Migdal:1956tc}
A.~B. Migdal,
\textit{``{Bremsstrahlung and pair production in condensed media at
  high-energies}''},
\doiref{10.1103/PhysRev.103.1811}{Phys.~Rev. \textbf{103}, 1811 (1956)}.

\bibitem{Landau:1953um}
L.~D. Landau \& I.~Pomeranchuk,
\textit{``{Limits of applicability of the theory of bremsstrahlung electrons
  and pair production at high-energies}''},
Dokl.~Akad.~Nauk~Ser.~Fiz. \textbf{92}, 535 (1953).

\bibitem{Braaten:1991we}
E.~Braaten \& M.~H. Thoma,
\textit{``{Energy loss of a heavy quark in the quark - gluon plasma}''},
\doiref{10.1103/PhysRevD.44.R2625}{Phys.~Rev.~D \textbf{44}, R2625 (1991)}.

\bibitem{THOMA1991491}
M.~H. Thoma \& M.~Gyulassy,
\textit{``Quark damping and energy loss in the high temperature QCD''},
\doiref{https://doi.org/10.1016/S0550-3213(05)80031-8}{Nuclear~Physics~B
  \textbf{351}, 491 (1991)}.

\bibitem{Djordjevic:2006tw}
M.~Djordjevic,
\textit{``{Collisional energy loss in a finite size QCD matter}''},
\doiref{10.1103/PhysRevC.74.064907}{Phys.~Rev.~C \textbf{74}, 064907 (2006)},
\normalsize{\texttt{\arxivref{nucl-th/0603066}{nucl-th/0603066}}}.

\bibitem{Barata:2022krd}
J.~a. Barata, A.~V. Sadofyev \& C.~A. Salgado,
\textit{``{Jet broadening in dense inhomogeneous matter}''},
\doiref{10.1103/PhysRevD.105.114010}{Phys.~Rev.~D \textbf{105}, 114010 (2022)},
\normalsize{\texttt{\arxivref{2202.08847}{arXiv:2202.08847}}}.

\bibitem{Andres:2022ndd}
C.~Andres, F.~Dominguez, A.~V. Sadofyev \& C.~A. Salgado,
\textit{``{Jet broadening in flowing matter: Resummation}''},
\doiref{10.1103/PhysRevD.106.074023}{Phys.~Rev.~D \textbf{106}, 074023 (2022)},
\normalsize{\texttt{\arxivref{2207.07141}{arXiv:2207.07141}}}.

\bibitem{Barata:2023qds}
J.~a. Barata, X.~Mayo~L\'opez, A.~V. Sadofyev \& C.~A. Salgado,
\textit{``{Medium induced gluon spectrum in dense inhomogeneous matter}''},
\doiref{10.1103/PhysRevD.108.034018}{Phys.~Rev.~D \textbf{108}, 034018 (2023)},
\normalsize{\texttt{\arxivref{2304.03712}{arXiv:2304.03712}}}.

\bibitem{Kuzmin:2023hko}
M.~V. Kuzmin, X.~Mayo~L\'opez, J.~Reiten \& A.~V. Sadofyev,
\textit{``{Jet quenching in anisotropic flowing matter}''},
\normalsize{\texttt{\arxivref{2309.00683}{arXiv:2309.00683}}}.

\bibitem{Kuzmin:2024smy}
M.~V. Kuzmin \& X.~Mayo~L\'opez,
\textit{``{Gluon radiation inside a flowing medium}''},
\normalsize{\texttt{\arxivref{2406.14628}{arXiv:2406.14628}}}.

\bibitem{Markert:2008jc}
C.~Markert, R.~Bellwied \& I.~Vitev,
\textit{``{Formation and decay of hadronic resonances in the QGP}''},
\doiref{10.1016/j.physletb.2008.08.073}{Phys.~Lett.~B \textbf{669}, 92 (2008)},
\normalsize{\texttt{\arxivref{0807.1509}{arXiv:0807.1509}}}.

\bibitem{Moffat:2021dji}
Jefferson Lab Angular Momentum (JAM) Collaboration, E.~Moffat, W.~Melnitchouk,
  T.~C. Rogers \& N.~Sato,
\textit{``{Simultaneous Monte~Carlo analysis of parton densities and
  fragmentation functions}''},
\doiref{10.1103/PhysRevD.104.016015}{Phys.~Rev.~D \textbf{104}, 016015 (2021)},
\normalsize{\texttt{\arxivref{2101.04664}{arXiv:2101.04664}}}.

\bibitem{Buckley:2014ana}
A.~Buckley, J.~Ferrando, S.~Lloyd, K.~Nordstr\"om, B.~Page, M.~R\"ufenacht,
  M.~Sch\"onherr \& G.~Watt,
\textit{``{LHAPDF6: parton density access in the LHC precision era}''},
\doiref{10.1140/epjc/s10052-015-3318-8}{Eur.~Phys.~J.~C \textbf{75}, 132
  (2015)},
\normalsize{\texttt{\arxivref{1412.7420}{arXiv:1412.7420}}}.

\bibitem{Greco:2003xt}
V.~Greco, C.~M. Ko \& P.~Levai,
\textit{``{Parton coalescence and anti-proton / pion anomaly at RHIC}''},
\doiref{10.1103/PhysRevLett.90.202302}{Phys.~Rev.~Lett. \textbf{90}, 202302
  (2003)},
\normalsize{\texttt{\arxivref{nucl-th/0301093}{nucl-th/0301093}}}.

\bibitem{PhysRevD.108.074013}
L.~S. Moriggi, E.~S. Rocha \& M.~V.~T. Machado,
\textit{``Study of the azimuthal asymmetry in heavy ion collisions combining
  initial state momentum orientation and final state collective effects''},
\doiref{10.1103/PhysRevD.108.074013}{Phys.~Rev.~D \textbf{108}, 074013 (2023)}.

\bibitem{Bilandzic:2012wva}
A.~Bilandzic,
\textit{``{Anisotropic flow measurements in ALICE at the large hadron
  collider}''}.

\bibitem{ALICE:2018yph}
ALICE Collaboration, S.~Acharya et~al.,
\textit{``{Anisotropic flow of identified particles in Pb-Pb collisions at $
  {\sqrt{s}}_{\mathrm{NN}}=5.02 $ TeV}''},
\doiref{10.1007/JHEP09(2018)006}{JHEP \textbf{1809}, 006 (2018)},
\normalsize{\texttt{\arxivref{1805.04390}{arXiv:1805.04390}}}.

\bibitem{CMS:2017xgk}
CMS Collaboration, A.~M. Sirunyan et~al.,
\textit{``{Azimuthal anisotropy of charged particles with transverse momentum
  up to 100 GeV/ c in PbPb collisions at $\sqrt {s}_{{NN}}$=5.02 TeV}''},
\doiref{10.1016/j.physletb.2017.11.041}{Phys.~Lett.~B \textbf{776}, 195
  (2018)},
\normalsize{\texttt{\arxivref{1702.00630}{arXiv:1702.00630}}}.

\bibitem{Ke:2024emw}
W.~Ke, J.~Terry \& I.~Vitev,
\textit{``{Anisotropic jet broadening and jet shape}''},
\normalsize{\texttt{\arxivref{2412.12250}{arXiv:2412.12250}}}.

\bibitem{Barata:2023zqg}
J.~a. Barata, J.~G. Milhano \& A.~V. Sadofyev,
\textit{``{Picturing QCD jets in anisotropic matter: from jet shapes to energy
  energy correlators}''},
\doiref{10.1140/epjc/s10052-024-12514-1}{Eur.~Phys.~J.~C \textbf{84}, 174
  (2024)},
\normalsize{\texttt{\arxivref{2308.01294}{arXiv:2308.01294}}}.

\bibitem{osg07}
R.~Pordes, D.~Petravick, B.~Kramer, D.~Olson, M.~Livny, A.~Roy, P.~Avery,
  K.~Blackburn, T.~Wenaus, F.~W{\"u}rthwein, I.~Foster, R.~Gardner, M.~Wilde,
  A.~Blatecky, J.~McGee \& R.~Quick,
\textit{``The open science grid''},
in \textit{``J. Phys. Conf. Ser.''},
p.~012057.
\bibitem{osg09}
I.~Sfiligoi, D.~C. Bradley, B.~Holzman, P.~Mhashilkar, S.~Padhi \&
  F.~Wurthwein,
\textit{``The pilot way to grid resources using glideinWMS''},
in \textit{``2009 WRI World Congress on Computer Science and Information
  Engineering''},
p.~428--432.
\bibitem{https://doi.org/10.21231/906p-4d78}
{OSG},
\textit{``\href{https://osg-htc.org/services/open\_science\_pool.html}{OSPool}''}.

\bibitem{https://doi.org/10.21231/0kvz-ve57}
{OSG},
\textit{``Open Science Data Federation''},
\href{https://osdf.osg-htc.org/}{\texttt{https://osdf.osg-htc.org/}}.

\end{thebibliography}
\end{document}